\documentclass[aps,twocolumn,groupedaddress,showpacs]{revtex4}
\usepackage{graphics}

\newcommand{\ra}{\rightarrow}
\newcommand{\al}{\alpha}
\newcommand{\bb}{\begin{eqnarray*}}
\newcommand{\ee}{\end{eqnarray*}}
\newcommand{\om}{\omega}
\newcommand{\Om}{\Omega}
\newcommand{\Z}{\langle}
\newcommand{\R}{\rangle}

\newcommand{\rr}{{\bf r}}
\newcommand{\vv}{{\bf v}}

\begin{document}

\title{Pseudo-forces in quantum mechanics}

\author{Pravabati Chingangbam\thanks{e-mail:prava@jamia.net}}
\author{Pankaj Sharan\thanks{e-mail:pankaj@jamia.net}}
\affiliation{Department of Physics, Jamia Millia Islamia  \\
	New Delhi-110 025, India    }


\begin{abstract}
Dynamical evolution is described as a parallel section on an 
infinite dimensional 
Hilbert bundle over the base manifold of all frames of reference. 
The parallel section is defined by an operator-valued connection whose 
components are the generators of the relativity 
group acting on the base manifold. In the case of Galilean 
transformations we show that the property that the curvature 
for the fundamental connection must be zero is 
just the Heisenberg equations of motion 
and the canonical commutation relation in geometric language. 
We then consider linear and circular accelerating frames 
and show that pseudo-forces must appear naturally in the 
Hamiltonian.

\end{abstract}


\maketitle

\section{Introduction}
Evolution of a state vector in quantum mechanics can be viewed
as a kind of parallel transport \cite{asorey}. 
There have been
suggestions to use the geometric language of vector bundles and parallel
transport in various situations in quantum mechanics\cite{prugo,drech,
gau,sarda,bozo}. These ideas 
are natural in the discussion of
the geometric or the Berry phase\cite{geo}. 

Despite these attempts to ``geometrize" quantum mechanics there seems to
be no common agreement in these approaches about the base space, or the
structure group, let alone the connection or the curvature. 
Moreover it is not clear
whether the extra mathematical machinary is justified by a new or clearer
physical insight. 

In this paper we give the {\em physical} reason why the bundle 
viewpoint is natural in quantum mechanics and illustrate it with 
application to accelerated frames. 

If a physical system is 
observed in various frames of reference, the states described
by them as vectors in their individual Hilbert spaces will form a
section in a vector bundle with the Hilbert space as the standard fiber
and the set of all frames of reference as the base manifold. There is no
canonical identification of the fibers and we need a ``connection", a 
notion of covariant derivative or that of parallel transport. 

We make use of the principle of relativity (all frames of reference are
equally suitable for description) to provide the notion of parallelism
and make the following assumption:
{\em States described by different frames of reference form a parallel
section.}

As each observer can apply an overall unitary operator on his
Hilbert space
and still obtain an equivalent description, we see that the structure group
should be the group of all unitary operators on the Hilbert 
space\cite{bohm}. 
Thus there is an underlying ``gauge freedom" 
which can be used to transform the natural parallel sections
into constant sections and do away with the need to use all Hilbert 
spaces at once. This is 
the case in standard quantum mechanics where a single Hilbert space is 
used by all observers. 

In this paper we  
develop our geometric picture and 
explicitly consider the case where Galilean group is the underlying 
relativity group.
We find that Heisenberg equations of motion and the canonical 
commutation relation are contained in a single condition:
that the 
fundamental connection is flat or that its curvature is zero.

Next we apply the geometric construction to accelerated frames 
and show that pseudo-force terms appear as expected. 
In the case of linearly accelerated frames we get a 
linear ``gravitational" potential implying that equivalence principle 
must hold in quantum mechanics. 
In contrast, in the conventional formalism equivalence principle is obtained 
by an artificial time dependent phase transformation of the 
wavefunction. 
In the case of rotating frames we show that 
both centrifugal and coriolis forces show up in the Hamiltonian. 
It is satisfactory to see that the coriolis force 
does not correspond to a potential because it does no work, being 
perpendicular to velocity, but naturally appears as 
a connection term added to the canonical momentum, much like the 
magnetic force. We are thus able to show that fibre bundles are the 
natural language in which to discuss quantum mechanical effects of gravity.

\section{Geometric setting}

\subsection{The bundle}

Consider a quantum mechanical system described by observers in different frames of 
reference. We assume that the set of all frames of reference forms a 
differentiable manifold. This is physically reasonable because frames 
of reference are related by symmetry transformations which form a 
group. This means that the frames can be labelled by coordinates $x$ on 
the group manifold.
A state of the system is 
described by a vector $\phi(x)$ in a Hilbert space ${\mathcal H}_x$ associated 
with the observer $x$. We, thus, have the ingredients of a vector 
bundle\cite{chern}. The base is a manifold $M$ with  
coordinates  $x$ and a Hilbert space at each 
point. To every possible state of the system is associated a  
section or a mapping $x \ra \phi(x)$ where $\phi(x)$ is the vector
describing the state of the system by observer $x$.
We assume there exist unitary operators $U(y,x)$ 
which connect the states $\phi(y) = U(y,x) \phi(x)$. 
These operators must satisfy consistency conditions
$$ U(z,y) U(y,x) = U(z,x);\qquad U(x,x)=1 $$

We must note that there is no canonical way of choosing states $\phi(x)$ 
to describe the system in the Hilbert space ${\mathcal H}_x$. 
If we were to apply 
a unitary operator to all vectors $\phi(x),\ \psi(x)$ etc. in 
${\mathcal H}_x$, the resulting states are equally 
well suited to describe the system  provided the observables  
acting in ${\mathcal H}_x$ are similarly changed. 
In other words, we assume the 
{\em structure} or {\em gauge group} acting on the fiber to be the 
group of all unitary operators. 

\subsection{The Connection}

Let us choose a complete orthonormal set $\phi_{\al}$  in the Hilbert 
space of some fixed observer, say at $x=0$, $\phi_{\al}\equiv\phi_{\al}(0)$. 
The sets ${\phi_{\al}(x)}=U(x,0)\phi_{\al}(0)$ 
then are complete orthonormal sets in all 
the other spaces ${\mathcal H}_x$. Any arbitrary section $\psi(x)$ can then be 
written as
$$ \psi(x)=\sum_{\al} c_{\al}(x)\phi_{\al}(x)  \eqno(1) $$
where $c_\alpha(x)$ are the complex coefficients of expansion.
Let $\Gamma$ be the set of all sections. They can be added pointwise.
$$ \big( \psi_1+\psi_2 \big)(x)=\psi_1(x)+\psi_2(x)  \eqno(2) $$
and multiplied with smooth functions 
$$(c\psi)(x)=c(x)\psi(x)            \eqno(3)  $$

Let $\Lambda\otimes\Gamma$ be the tensor product of the space $\Lambda$ 
of all 1-forms on the base $M$ and $\Gamma$. A connection on this 
bundle is a mapping $D:\Gamma \ra \Lambda\otimes\Gamma $ such that
\bb
D(\psi_1+\psi_2) &=& D\psi_1+D\psi_2       \qquad {\rm and} \\
D(c\psi)         &=& cD\psi +dc.\psi
\ee
$$ \eqno(4)  $$
If $\phi_n(x)$ is a basis in $\Gamma$ we can express $D(\phi_n)$ in 
terms of the basis $dx^\mu \otimes \phi_m$ in $\Lambda\otimes\Gamma$ as
$$ (D\phi_n)(x) = \phi_m(x) \Gamma_{\mu m n} dx^\mu  \eqno(5) $$
where coefficients $ \Gamma_{\mu mn}(x)$ are the 
Christoffel symbols with respect to the basis  $dx^\mu\otimes\phi_m$. 
We write this equation as 
$$(D\phi_n)(x) = \phi_m\om^{\phi}_{mn}(x)  \eqno(6)$$
where the complex matrix $\omega_{mn}$ can be obtained by taking inner 
product with  $\phi_m$ in Eq.(5).
$$ \om^{\phi}_{mn} = \big( \phi_m,D\phi_n\big) \eqno(7)$$
This matrix of one-forms is called the {\em connection 
matrix}. 
We require $D$ to satisfy Leibniz rule 
$$ D(\phi,\psi) = (D\phi,\psi) + (\phi,D\psi) = d(\phi,\psi) \eqno(8) $$
which when applied to $\delta_{mn} = (\phi_m,\phi_n)$ shows
that $\om^{\phi}$ is an anti-hermitian matrix.

\noindent Under a change of basis 
$$ \chi_n(x) = U(x) \phi_n(x)     \eqno(9)   $$
we have
\bb
\chi_n(x)  &=& \phi_m(x)\big( \phi_m(x), U(x) \phi_n(x)\big)  \\
           &=& \phi_m(x)U_{mn}(x)
\ee	   
$$ \eqno(10)     $$
\noindent Thus, omitting the base point $x$ for simplicity of notation
\bb
(D\chi_n)(x)  &=& D\big(\phi_s U_{sn}\big) \\
              &=& \phi_r \om^{\phi}_{rs}U_{sn} + \phi_s dU_{sn} \\
              &=& \chi_m\om^{\chi}_{mn} = \phi_r U_{rm} \om^{\chi}_{mn}                  	       	     
\ee
$$ \eqno(11)   $$
Or,
$$ \om^{\chi}_{mn} = U^{-1}_{mr} \om^{\phi}_{rs}U_{sn} 
                       + U^{-1}_{mr}dU_{rn}          \eqno(12)  $$
Omitting matrix indices, we have		       
$$ \om^{\chi} = U^{-1} \om^{\phi}U + U^{-1}dU    \eqno(13)  $$
The curvature 2-form for the connection is given by
$$ \Om^{\phi} = d\om^{\phi} + \om^{\phi} \wedge \om^{\phi}  \eqno(14) $$
which transforms as 
$$\Om^\chi = U^{-1}\Om^{\phi}U  \eqno(15)   $$
One may also note the Bianchi identity
$$ d\Om = \Om \wedge \om - \om\wedge\Om  \eqno(16)  $$

\section{Parallel section and the fundamental connection}

We now make the fundamental assumption that a system observed by
different observers is represented by parallel sections.
Let $ \phi_m(x)$ be a family of parallel sections, that is
$$ (D\phi)(x)=0, \qquad\qquad {\rm for\,\,all}\ m \eqno(17)  $$
This implies
$$ \om^{\phi}_{mn}(x) =0 \eqno(18)  $$
everywhere.

We shall now see how does the connection matrix look like with respect to 
the basis of constant sections. The advantage of using constant 
sections is that one can give up the bundle picture altogether and 
identify all Hilbert spaces together to work in one common space.
The constant section physically means that the state is represented by the same 
constant vector by all observers. {\em This is the most general definition 
of the Heisenberg picture}.

To get constant sections we use the fact that parallel sections 
are constructed by applying transformation $U(x,0)$ on $\phi(0)$ 
for all $x$. 
$$ \phi_m(x) = U(x,0)\phi_m(0) =U(x)\phi_m(0)   \eqno(19)  $$
We can choose $\phi_m(0)$ as the new basis
\bb
\chi_m(x) \equiv \phi_m(0) &=& U_x^{-1} \phi_m(x)\\
                           &=& \phi_r(x) \big(\phi_r(x), 
                                U^{-1}(x)\phi_m(x)\big) \\
			   &=& \phi_r(x)U^{-1}_{rm}
\ee
$$  \eqno(20)    $$
Then
$$ \om^{\chi}=UdU^{-1}    \eqno(21)   $$
which, as expected, is pure gauge.

\section{Galilean frames}

Let us consider a particle of mass $m$ in one space dimension.
We use units where $c=\hbar=1$. 
We consider the basis of  sharp momentum states $|k\R$ such that 
$$ P|k\R = k|k\R           \eqno(22) $$
and
$$ \Z k'|k\R = \delta(k-k')            \eqno(23)$$
The time and space translations are given by the operators $U_{\tau}$ and 
$U_{\zeta}$ respectively,
$$ U_\tau |k\R = e^{-iH\tau}|k\R =e^{-i{{k^2}\over{2m}}}|k\R   \eqno(24)  $$
$$ U_\zeta |k\R = e^{iP\zeta}|k\R = e^{ik\zeta}|k\R   \eqno(25)  $$
The boosts act as 
$$ U_\eta |k\R = |k-m\eta\R     \eqno(26)      $$
given by
$$ U_\eta = e^{-iK\eta}     \eqno(27)    $$
where $K$ is the boost generator.
The lie algebra of the Galilean group is 
$$ [P,H]=0,\hskip1cm [K,H]=iP  $$
$$ [K,P]=imI                        \eqno(28)  $$ 

The algebra is not closed. This is because unitary representation of 
the Galilean group in $\mathcal H$ is projective. The position operator 
$X$ is related to $K$\cite{ali} as
$$ K= mX    \eqno(29)  $$
and it acts on the states $|k\R$  as
$$ \Z k|X = i{{\partial}\over{\partial k}} \Z k|  \eqno(30) $$
Parallel sections can be constructed using $U_\tau$, $U_\zeta$ and 
$U_\eta$ in a 
variety of ways. We choose the following convention which corresponds 
to the transformations
\bb
x' &=& x-\eta t - \zeta   \\
t' &=& t-\tau  \\
v' &=& v - \eta
\ee
$$\eqno(31)   $$
between frame $S$ and $S'$. If we take the standard frame at $x=0,\ 
t=0,\ v=0$ then
\bb
\phi(x',t',v')   &=&  \phi(-\zeta,-\tau,-\eta)   \\
                 &=&  U_{\tau}U_{\zeta}U_{\eta} \phi(0,0,0)   
\ee
$$                                      \eqno(32)      $$
We rename coordinates
$$ \phi(x,t,v) =  U_{-t}U_{-x}U_{-v} \phi(0,0,0)   \eqno(33)  $$ 
            
and get for the basis of constant sections 
\bb
\om^{\chi}  &=& U_{-t}U_{-x}U_{-v} d\Big(U_{-t}U_{-x}U_{-v}\Big)^{-1}  \\
            &=& i\Big( -Hdt + Pdx +X(t)mdv -mxdv \Big)
\ee	    
$$  \eqno(34)      $$
The curvature is zero, as it should be for a pure gauge connection. But 
it is worth seeing explicitly. 
$$ \Om^{\chi} = d\om + \om \wedge\om = 0       \eqno(35)      $$
This implies the following equations
$$ i{\dot P}  = [P,H]   \eqno(36)   $$
$$ i{\dot X}  = [X,H]   \eqno(37)    $$
$$  [X,P] = i     \eqno(38)     $$
Equations (34) and (35) are just the Heisenberg equations of motion for 
operators $P$ and $X$ while the third is the canonical commutation 
relation for $X$ and $P$. 

One may argue that these equations are just reproductions of the
algebra. Indeed the algebra is used in the calculation of the 
curvature. What is new is that in this differential geometric language 
all the information is contained in a single zero curvature equation.

\section{Accelerated frames and pseudo forces}

Acceleration implies changing from one Galilean frame to another after 
every infinitesimal amount of time. This can be seen as a curve on 
the base manifold parametrized by time. We assume that an
observer in the accelerating frame uses the same Hilbert space to 
describe a physical system as 
the observer at the base manifold point with 
which it coincides at each instant $t$. 
Moreover they assign the same state to the system\cite{bell}.

\subsection{Linearly accelerated frame and equivalence principle}
The question of whether the principle of equivalence in classical 
mechanics also holds in quantum mechanics was discussed by C.J. 
Eliezer and P.G. Leach\cite{eliezer}. They studied the transformation 
of the Schr\"odinger equation under a change from an inertial frame of 
reference $S$ to a uniformly accelerating one $S'$. 
Their argument goes as follows. Let
$$ x' = x+ {{1}\over{2}} gt^2, \qquad t'=t  \eqno(39)   $$
be the change of coordinates to an accelerated frame. Then
he equivalence 
principle holds provided the phase of the wavefunction of the system is 
redefined by a time dependent expression. This means that the Schr\"odinger 
equation in the frame $S$
$$ i {{d\psi}\over{dt}} = -{{1}\over{2m}} \nabla^2\psi  
                                                    \eqno(40)   $$
transforms to the equation for a particle moving in a uniform field
$$ i{{d\psi'}\over{dt'}} = -{{1}\over{2m}} \nabla'^2\psi'  
                                  - mgx' \psi'              \eqno(41)   $$
with the redefinition of the phase of $\psi$ given by
$$ \psi'(x') = exp\Big({{img}\over{\hbar}}(x't'-{{1}\over{6}}gt'^3\Big) \psi(x)   
                                                   \eqno(42)   $$
The phase factor has been chosen precisely to obtain equivalence 
principle. There is no explanation put forward for this factor.

In our formalism we find that the equivalence principle must hold in quantum 
mechanics in a straightforward manner. There is no need for any other 
condition such as the redefinition of the wavefunction by a time-dependent 
phase factor, like the one seen above. 

Consider an observer in a linearly accelerated frame of reference.
The linear acceleration corresponds to a curve on the base  
manifold parametrized by $t$ and given 
by  
$$ x  = {{1}\over{2}}gt^2   \eqno(43)   $$
$$ v  = gt   \eqno(44)     $$

The parallel section is again specified by
$$ |t,x,v;k\R  \equiv  U_vU_xU_t |k\R    \eqno(45)       $$
The rate of change of the vector along the curve should give the 
Hamiltonian for the accelerated observer.
We get
\bb
i{{d}\over{dt}}|t,x,v;k\R &=& i{{d}\over{dt}} \Big(U_tU_xU_v|k\R\Big)\\
                          &=& \Big( {{(k-mv)^2}\over{2m}} 
			       + (k-mv)t{{dv}\over{dt}}
                               - (k-mv){{dx}\over{dt}}\\
			  &{}& - mx{{dv}\over{dt}} - mgx
			       - i{{dv}\over{dt}}
			           {{\partial}\over{\partial v}}\Big)
				   |t,x,v;k\R  \\
                          &=& \Big( {{P(v)^2}\over{2m}} - X(x)mg 
			        \Big)|t,x,v;k\R 
\ee				
$$\eqno(46) $$
where $P(v)=k-mv$ and $X(x)=X+x$.
Thus, the system ``sees" an extra potential $X(x)mg$ which is the expected 
linear ``gravitational" potential term. This is manifestation of the equivalence principle in 
quantum mechanics.  

The validity of the equivalence principle in the quantum regime has 
been experimentally tested in some beautiful experiments done with 
neutron interference\cite{cow}.
		                      	       	     
\subsection{Rotating frame, coriolis and centrifugal forces}

Consider a frame of reference $S'$ which is rotating with constant angular 
velocity $\omega$ and radius $r$ about the origin of coordinates  
on the $xy$ plane of a frame $S$. 
The two frames are related as follows: we wait for time $t$,
translate by ${\rr}$ direction, rotate by angle 
$\theta=\omega t$ and finally give a boost in the $y'$ direction by 
velocity $v$.

\begin{figure}
\resizebox{!}{3in}{\includegraphics{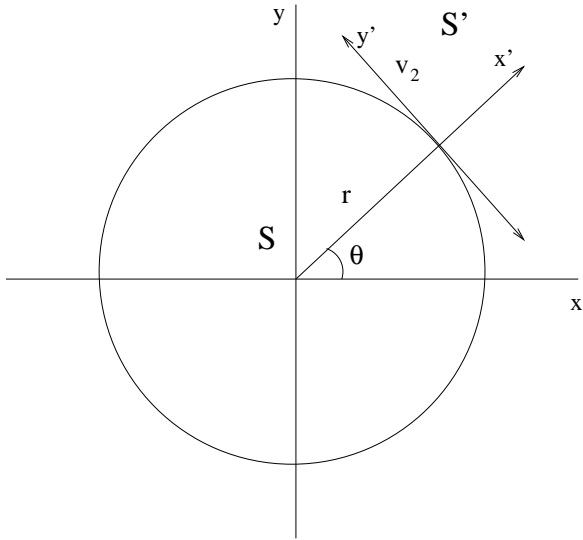}}
\caption{ $S'$ is a frame which is rotating with angular 
velocity $\omega$ about origin of frame $S$ with radius $r$.}
\end{figure}

The parallel section is given by
\bb
U &=& U_vU_\theta U_{\rr} U_t   \\
  &=& e^{-i{\vec X}_2 mv} e^{iJ\theta}
      e^{iP.{\rr}} e^{-i{{P^2}\over{2m}}t}
\ee
$$\eqno(47)$$
where $J=X_1P_2-X_2P_1$ is the angular momentum operator. 
The curve on the base manifold parametrized by 
$t$ is 
\bb
{\rr} &=& (r\,{\rm cos} \,\theta, r\,{\rm sin}\,\theta)  \\
{\vv} &=& (-r\omega\,{\rm sin}\,\theta, r\omega\,{\rm cos}\,\theta)
\ee
$$ \eqno(48)   $$

The Hamiltonian $H$ as seen by an observer 
in the rotating frame is given by the rate of 
change of the vectors specified along the curve on the base 
manifold.
\bb
H &=& i{{dU}\over{dt}}U^{-1}  \\
  &=& {{1}\over{2m}}\Big( P_1^2 + (P_2+m\omega r)^2 \Big)
      -\omega r(P_2 +m\omega r)  \\
  &{}&      -\omega\big( J + m\omega r\big) 
\ee
$$\eqno(49)$$
or
\bb
H &=&{{1}\over{2m}}\Big( (P_1 + m\omega X_2)^2 + (P_2-m\omega X_1)^2 
             \Big)\\
  &{}&  -{{1}\over{2}}m\omega^2 \big( (X_1+r)^2 + X_2^2 
\big)    
\ee
$$ \eqno(50)  $$

Thus the expected centrifugal and coriolis forces appear in the 
Hamiltonian. Since 
coriolis force does no work it cannot appear as an explicit potential 
term. Rather it appears as a connection in the canonical 
momentum.

\section{Discussion}

The bundle viewpoint is hinted in Dirac's work as 
early as 1932. In a most influential paper {\em The Lagrangian in 
quantum mechanics}\cite{dirac} Dirac puts forth 
the following argument:
Let $q(t)$ be  a complete set of commuting observables in the Heisenberg
picture. The set of eigenvalues $q'$ at each $t$ forms a manifold $M$
giving rise to ``spacetime" $B\equiv M\times T$ where $T$ 
represents the time axis.

Evolution is determined by the moving basis $\Z q',t| $ at each $ (q',t) $. 
This can be interpreted as a section from the base $B$ into a Hilbert 
space.
Let  $ c:\tau\ra (q'(\tau),t(\tau)) $ be a curve in $B$.
Then the change of basis vectors $\Z q',t| $ is given by
$$ -i{\partial\over{\partial q'}}\Z q',t| = \Z q',t|\,P(t)    $$
$$ i{\partial\over{\partial t}}\Z q',t| = \Z q',t|\,H        $$
where $P(t)=e^{iHt}P(0)e^{-iHt}$

Thus the change of a basis vector along the curve is

$$ d\Z q',t|=i \Z q',t|\,dS $$
$$ dS=P(t)dq'-Hdt      $$

Thus in the bundle formalism Dirac's Lagrangian can be seen as an operator 
-valued 1-form on the Hilbert vector bundle whose base manifold is spanned 
by the eigenvalues $q'$ of a complete set of commuting operators 
$q(t)$ specified at all times and the 
standard fibre is Hilbert space. The components of this 1-form are 
just the Hamiltonian and momentum operators. 
If the section 
$(q',t)\to \Z q',t| $ 
is 
assumed to be parallel then evolution can be seen as parallel 
transport. This is the theme on which our present work is based.

Asorey et al. \cite{asorey} consider a Hilbert bundle with positive 
time axis ${\mathcal R}^+$ as the base manifold. 
Another viewpoint is that of Prugovecki\cite{prugo} 
and Drechsler and Tuckey\cite{drech} whose bundle is the associated vector bundle 
for the principle bundle with Poincare group as structure group over 
spacetime base manifold. The Hilbert space considered by them is the 
space of square integrable functions over phase space of space 
coordinates and the mass hyperboloid ($p^2=m^2,\ p_0>0$). This approach 
allows them to consider parallel transport over curved spaces with 
possible applications to quantum gravity.

Dirk Graudenz\cite{gau} also has a Hilbert bundle with 
spacetime base. Our approach agrees with that of Graudenz in that 
description of a physical system is always description by one observer. 
Yet another construction is given by G. Sardanashvily\cite{sarda} 
who consider a $C^*$-algebra at each point of the time axis ${\mathcal R}$.

Our geometric construction is different from others in the 
literature. For us the base manifold consists of all frames of 
reference. This means actually having a group of symmetry 
transformations as the base manifold with a frame of reference 
associated with each point on it. We have considered the case of
Galilean group which makes the application specific to quantum 
mechanics. 

Our objective here is to present a new geometric viewpoint 
from which implies the validity of the equivalence principle 
in quantum mechanics. We have demonstrated this both for linearly 
accelerating and rotating frames.

\section*{Acknowledgements}{One of us (P. Chingangbam) 
acknowledges financial support 
from Council for Scientific and Industrial Research, India, under 
grant no.9/466(29)/96-EMR-I. We thank Tabish Qureshi for useful 
discussions.}

\end{document}